\documentclass[prd,notitlepage,longbibliography,nofootinbib,superscriptaddress,onecolumn,preprintnumbers]{revtex4-2}
\usepackage[utf8]{inputenc}

\usepackage{bm}
\usepackage{comment} 
\usepackage[colorlinks=true,urlcolor=blue,anchorcolor=black,citecolor=blue,linkcolor=red,filecolor=black,menucolor=black,pagecolor=black,linktocpage=true,pdfproducer=medialab,pdfa=true]{hyperref}
\usepackage{graphicx}
\usepackage{amsmath,latexsym,amssymb,mathrsfs,ascmac,mathtools}
\usepackage{multirow}
\usepackage{braket}
\usepackage{placeins}
\usepackage{float}
\allowdisplaybreaks
\begin{document}

\title{EFT Corrections to Photon Propagation and Gravitational Lensing in an Ellis–Bronnikov Wormhole}

\author{Takamasa Kanai}
\email{kanai@kochi-ct.ac.jp}

\affiliation{Department of Social Design Engineering,
National Institute of Technology (KOSEN), Kochi College,
200-1 Monobe Otsu, Nankoku, Kochi, 783-8508, Japan}

\begin{abstract}
We investigate strong gravitational lensing by an Ellis–Bronnikov (EB) wormhole in the presence of effective field theory (EFT) corrections to photon propagation. We derive the modified photon propagation law induced by non-minimal couplings between the electromagnetic field and spacetime curvature, which leads to polarization-dependent photon trajectories in the EB wormhole spacetime. Using the strong deflection limit, we derive the corrections to the photon sphere and the logarithmically divergent part of the deflection angle. Although the EFT corrections are parametrically small, their contributions can become appreciable near the photon sphere. In particular, the $R_{\mu\nu}F^{\mu\rho}F^{\nu}{}_{\rho}$ interaction, which is nonvanishing for the EB wormhole but absent in the Schwarzschild spacetime, contributes to the strong-deflection coefficients. This may provide a distinctive lensing signature for observationally distinguishing wormholes from black holes.
\end{abstract}
\maketitle

\section{Introduction}

Gravitational phenomena in the strong-field regime provide a unique window into the nature of gravity and the structure of compact objects beyond the weak-field approximation. In particular, the propagation of light near unstable photon orbits gives rise to characteristic observables such as black hole shadows \cite{Falcke:1999pj,EventHorizonTelescope:2019dse,EventHorizonTelescope:2022wkp,Vagnozzi:2022moj} and gravitational lensing \cite{Virbhadra:1999nm,Bozza:2001xd,Bozza:2002zj,Gibbons:2008rj}. These observables are especially interesting for testing exotic compact objects, since their strong-field gravitational signatures can differ from those of black holes.

Among these probes, strong gravitational lensing provides a particularly sensitive probe of the geometry in the vicinity of an unstable photon orbit. In the strong deflection limit, the deflection angle develops a logarithmic divergence as the closest approach radius approaches the photon-sphere radius. This behavior is captured by the formalism developed by Bozza \cite{Bozza:2002zj,Bozza:2002af,Bozza:2003cp}, in which the deflection angle is characterized by a logarithmically divergent term and a finite regular contribution. The coefficients of this expansion are determined by the local properties of the effective geometry near the photon sphere, making strong lensing particularly sensitive to small deviations from the underlying spacetime geometry.

In the framework of effective field theory (EFT) \cite{Weinberg:1978kz,Donoghue:1993eb,Donoghue:1994dn,Burgess:2003jk}, higher-derivative interactions can modify not only the background geometry but also the propagation of electromagnetic waves. While gravitational lensing is conventionally described by null geodesics of the background metric, non-minimal couplings between the electromagnetic field and spacetime curvature can modify the propagation of photons themselves. In particular, such interactions alter the photon dispersion relation, so that the resulting characteristic surfaces need not coincide with the null cones of the background metric. Moreover, these surfaces can depend on the polarization of the photon, leading to gravitational birefringence \cite{Scharnhorst:1990sr,Barton:1989dq,Barton:1992pq,Latorre:1994cv,Dittrich:1998fy,DeLorenci:2000yh,Drummond:1979pp,Daniels:1993yi,Shore:1995fz,Daniels:1995yw,Shore:2007um,Cho:1997vg,Izumi:2014loa,Reall:2014pwa,Allahyari:2019jqz,Cao:2021sty,Reall:2021voz,Davies:2021frz,Fu:2025oxr}. Consequently, the relevant photon sphere and strong-deflection coefficients are determined not only by the background spacetime but also by the effective propagation of light.

This distinction is particularly important for wormhole spacetimes, whose curvature structure can differ qualitatively from that of vacuum black holes. Previous studies have investigated the possibility of distinguishing wormholes from black holes through their characteristic shadows, showing that differences in photon trajectories near unstable photon orbits can leave observable imprints on the shadow structure \cite{Takahashi:2026hfp}. These studies demonstrate that photon propagation in the strong-field region can provide a useful means of probing the nature of compact objects and distinguishing exotic objects from black holes.

In this work, we focus on the Ellis-Bronnikov (EB) wormhole \cite{Ellis:1973yv,Bronnikov:1973fh}, which is not a vacuum solution of the Einstein equations and possesses nonvanishing Ricci curvature. This makes the EB wormhole an especially useful setting for studying curvature-dependent photon interactions that are absent or trivial in Ricci-flat black hole spacetimes. In particular, the nonvanishing Ricci curvature allows interactions involving the Ricci tensor to contribute in addition to Weyl curvature couplings. While curvature-photon couplings and their effects on photon propagation have been studied in various curved spacetimes, their implications for wormhole geometries with nonvanishing Ricci curvature and their consequences for strong-deflection lensing warrant further investigation.

In this work, we derive the modified photon dispersion relations in the eikonal approximation and construct the corresponding effective optical metrics for the two physical polarization modes in the EFT-corrected EB wormhole spacetime. The resulting photon trajectories exhibit polarization dependence, providing a gravitational analogue of birefringence. We then apply the strong deflection limit formalism to these curvature-coupled photons and derive the corrections to the photon-sphere radius and the strong-deflection coefficients. In particular, we find that the EFT corrections contribute to the logarithmically divergent part of the deflection angle. Although these corrections are perturbatively small, their effects can become appreciable in the strong deflection regime, where photons spend a long time in the vicinity of the unstable photon orbit. This provides a potentially sensitive way to probe curvature-photon interactions through strong gravitational lensing.

A particularly distinctive feature of the EB wormhole is the contribution from the interaction $R_{\mu\nu}F^{\mu\rho}F^{\nu}_{\ \rho}$, which is nonvanishing in the wormhole geometry but vanishes in the Schwarzschild spacetime because $R_{\mu\nu}=0$. Its contribution to the strong-deflection coefficients therefore provides a characteristic difference between the two spacetimes. Rather than producing a universal modification of the lensing observables, the EFT corrections retain information about the curvature structure of the underlying spacetime. This may leave a characteristic imprint on strong gravitational lensing and provide a possible avenue for distinguishing wormhole geometries from vacuum black holes.

We further investigate the photon time delay as a complementary observable of the modified photon propagation. Unlike the deflection angle, the time delay does not exhibit the same logarithmically divergent behavior in the strong-deflection limit, and its differences between wormhole and black-hole spacetimes may therefore be less pronounced. Nevertheless, the time delay provides an independent probe of the EFT corrections and can complement the lensing observables. By combining these observables, we explore how curvature-photon couplings can imprint the non-vacuum nature of the EB wormhole on photon propagation and assess the prospects for identifying characteristic signatures of EFT-corrected wormhole spacetimes.

In Sec.~\ref{sec:EB}, we introduce the effective action for the EB wormhole and derive the modified photon propagation law induced by the curvature-photon couplings, together with the corresponding effective metric. In Sec.~\ref{sec:strong lens}, we investigate gravitational lensing in the strong-deflection limit and derive the effects of the EFT corrections on the photon sphere and the deflection angle. We further discuss the characteristic features of the lensing observables in the EB wormhole spacetime and highlight their differences from those expected in the Schwarzschild black hole spacetime. In Sec.~\ref{sec:time delay}, we study the photon time delay in the strong-deflection regime and examine its sensitivity to the EFT corrections. Finally, in Sec.~\ref{sec:conclusion}, we summarize our results and discuss their implications and future prospects.

In this paper, we set the Newton constant and the speed of light equal to unity.

\section{Photon Propagation and Effective Geometry in EFT-Corrected EB Wormhole}
\label{sec:EB}

Gravitational lensing by compact objects and exotic spacetimes provides a powerful probe of strong-field gravity and the underlying spacetime geometry~\cite{Einstein:1936llh,Virbhadra:1999nm,Bozza:2001xd,Bozza:2002zj,Gibbons:2008rj}. Beyond the geometry of the background spacetime, the propagation of light can also be modified by non-minimal couplings between the electromagnetic field and spacetime curvature. Such curvature couplings arise naturally from quantum corrections to electrodynamics in curved spacetime~\cite{Drummond:1979pp} and can lead to polarization-dependent photon propagation, giving rise to a gravitational analogue of birefringence~\cite{Chen:2015cpa,Lu:2016gsf}. Among the possible curvature couplings, the interaction between the electromagnetic field and the Weyl tensor has been extensively studied, including its effects on photon propagation and gravitational lensing in black-hole spacetimes~\cite{Drummond:1979pp,Chen:2015cpa,Lu:2016gsf,Chen:2016hil}.

In this work, we consider a broader class of curvature couplings that includes not only the Weyl tensor but also the Ricci tensor. We investigate the propagation of photons in an Ellis-Bronnikov (EB) wormhole spacetime and study how these curvature couplings modify the photon trajectories and the resulting gravitational lensing. In contrast to the minimally coupled case, the curvature-photon interactions lead to polarization-dependent light-cone conditions, and hence to distinct effective optical metrics for the two polarization modes. We derive these effective metrics from the photon dispersion relation in the eikonal approximation and use them to analyze the gravitational lensing of curvature-coupled photons in the EB wormhole background. The resulting framework allows us to clarify how higher-curvature interactions and the nontrivial wormhole geometry jointly affect the propagation of light.

\subsection{Equation of motion for a higher curvature-coupled photon}

The starting point of our analysis is the effective action for an electromagnetic field and a scalar field in a curved spacetime, including higher-curvature and curvature-matter interactions. In particular, we consider the action
\begin{align}
S=\int d^{4}x\sqrt{-g}\Biggl[&\frac{R}{16\pi G}+8\pi\nabla_{\mu}\Phi\nabla^{\mu}\Phi-\frac{1}{4}F_{\mu\nu}F^{\mu\nu}+\alpha_1 R^2+\alpha_2R_{\mu\nu}R^{\mu\nu}+\alpha_3R_{\mu\nu\rho\sigma}R^{\mu\nu\rho\sigma}\notag\\
&+\beta_1RF_{\mu\nu}F^{\mu\nu}+\beta_2R_{\mu\nu}F^{\mu\rho}F^\nu_{\ \rho}+\beta_3\, C^{\mu\nu\rho\sigma}F_{\mu\nu}F_{\rho\sigma}\notag\\
&+\gamma_1\Box\Phi F_{\mu\nu}F^{\mu\nu}+\gamma_2\nabla_\mu\Phi\nabla^\mu\Phi F_{\nu\rho}F^{\nu\rho}+\gamma_3\nabla_\mu\Phi\nabla_\nu\Phi F^{\mu\rho}F^{\nu}_{\ \rho}\notag\\
&+\lambda_1(\Box\Phi)^2+\lambda_2\Box\Phi\nabla_\mu\Phi\nabla^\mu\Phi+\lambda_3(\nabla_\mu\Phi\nabla^\mu\Phi)^2+\xi_1(F_{\mu\nu}F^{\mu\nu})^2+\xi_2F_{\mu\nu}F^{\nu}_{\ \rho}F^{\rho}_{\ \sigma}F^{\sigma\mu}\notag\\
&+\kappa_1\Box\Phi R+\kappa_2\nabla_\mu\Phi\nabla^\mu\Phi R+\kappa_3\nabla_\mu\Phi\nabla_\nu\Phi R^{\mu\nu}\Biggr],
\label{eq:action}
\end{align}
where $F_{\mu\nu}$ denotes the Maxwell field strength, $\Phi$ is a scalar field, and $\alpha_i$, $\beta_i$, $\gamma_i$, $\lambda_i$, $\xi_i$, and $\kappa_i$ are EFT coupling constants with dimensions of length squared. Among these terms, the couplings between the electromagnetic field and the Ricci tensor and Weyl tensor, parametrized by $\beta_2$ and $\beta_3$, respectively, play a central role in modifying the photon propagation law in the curved spacetime considered below. In particular, the Ricci-photon coupling parametrized by $\beta_2$ provides a distinctive contribution in the EB wormhole spacetime due to its nonvanishing Ricci curvature. The Ricci curvature encodes the contribution of matter and non-gravitational sources to the spacetime curvature, whereas the Weyl curvature represents the gravitational degrees of freedom associated with the free gravitational field. The Weyl-photon coupling parametrized by $\beta_3$ therefore modifies photon propagation through the Weyl curvature and provides a probe of the gravitational degrees of freedom of the background spacetime. The Weyl tensor are defined as
\begin{equation}
C_{\mu\nu\rho\sigma}=R_{\mu\nu\rho\sigma}-\big(g_{\mu[\rho}R_{\sigma]\nu}-g_{\nu[\rho}R_{\sigma]\mu}\big)+\tfrac{1}{3}R\,g_{\mu[\rho}g_{\sigma]\nu}.
\label{eq:weyl-def}
\end{equation}

We introduce a perturbative redefined field as follows:
\begin{align}
g_{\mu\nu}=\hat{g}_{\mu\nu}+\delta g_{\mu\nu},
\end{align}
with
\begin{align}
\delta g_{\mu\nu}=&a_1 Rg_{\mu\nu}+a_2R_{\mu\nu}+a_3F_{\nu\rho}F^{\nu\rho}g_{\mu\nu}+a_4F^{\mu\rho}F^{\nu}_{\ \rho}+a_5\Box\Phi g_{\mu\nu}\notag\\
&+a_6\nabla_\mu\Phi\nabla^\mu\Phi g_{\mu\nu}+a_7\nabla_\mu\Phi\nabla_\nu\Phi.
\end{align}
By appropriately choosing the coefficients $a_1,\ldots,a_7$, the above field redefinition can be used to eliminate seven operators from the action. Since our primary focus is on gravitational lensing effects arising from modifications to photon propagation, we restrict our attention to operators that directly affect the propagation of light. In particular, we do not consider terms such as $(\partial_\mu\Phi\partial^\mu\Phi)^2$, which can induce a perturbative correction to the metric at first order but do not modify the photon propagation at the order of interest. On the other hand, the curvature-photon coupling terms retained below do not modify the background metric at first order, since the background electromagnetic field is assumed to vanish, $F_{\mu\nu}=0$. Nevertheless, these terms contribute to the photon propagation at leading order in the eikonal approximation and are therefore essential for studying the gravitational lensing of photons. Accordingly, we consider the following action as the relevant effective theory for our analysis:
\begin{align}
S=\int d^{4}x\sqrt{-g}\Biggl[&\frac{R}{16\pi G}+8\pi\nabla_{\mu}\Phi\nabla^{\mu}\Phi-\frac{1}{4}F_{\mu\nu}F^{\mu\nu}+\alpha RF_{\mu\nu}F^{\mu\nu}+\beta R_{\mu\nu}F^{\mu\rho}F^\nu_{\ \rho}+\gamma C^{\mu\nu\rho\sigma}F_{\mu\nu}F_{\rho\sigma}\Biggr].
\label{eq:action}
\end{align}

Varying the action~\eqref{eq:action} gives a modified Maxwell
equation,
\begin{align}
\nabla_{\mu}\!\left(F^{\mu\nu}-4\alpha RF^{\mu\nu}-2\beta R_{\mu\rho}F^{\rho\nu}-2\beta R^{\nu\rho}F^{\mu}_{\ \rho}-4\gamma\,C^{\mu\nu\rho\sigma}F_{\rho\sigma}\right)=0 .
\label{eq:maxwell-mod}
\end{align}
To investigate photon propagation in the geometric-optics regime, we adopt the eikonal approximation, where $a^\mu$ denotes the polarization vector  satisfying the transversality condition $k_\mu a^\mu=0$. We then substitute the eikonal ansatz into the modified Maxwell equation and retain the leading-order terms in the eikonal expansion, yielding the following dispersion relation for the photon:
\begin{align}
(1-4\alpha R)k_{\mu}k^{\mu}a^{\nu}-2\beta R_{\mu\rho}k^\mu k^\rho a^\nu+2\beta R_{\mu\rho}k^\mu a^\rho k^\nu-2\beta R^{\rho\nu}a_\rho k_{\mu}k^{\mu}+8\gamma\,C^{\mu\nu\rho\sigma}k_{\sigma}k_{\mu}a_{\rho}=0.
\label{eq:eom-photon}
\end{align}
The term quadratic in the Ricci scalar and the fourth term in the above equation are both proportional to $k_{\mu}k^{\mu}$. Since these terms vanish on the leading-order null cone, they do not contribute to the perturbative correction to the effective metric considered in this work. Therefore, we omit these terms in the following analysis.

\subsection{EB wormhole background and the Weyl tensor in an orthonormal frame}

The phantom scalar field and the background metric of the EB wormhole are given by
\begin{align}
\Phi&=\sqrt{\dfrac{M^2+\ell^2}{4\pi\ell^2}}\left[\arctan\left(\frac{r}{\ell}\right)-\frac{\pi}{2}\right],\\
ds^{2}&=-e^{h}dt^{2}+e^{-h}[dr^2+(r^2+\ell^2)(d\theta^{2}+\sin^2\theta d\phi^2)],
\label{eq:EB-metric}
\end{align}
with
\begin{equation}
h(r)\equiv\dfrac{2M}{\ell}\left[\arctan\left(\frac{r}{\ell}\right)-\frac{\pi}{2}\right],
\label{eq:EB-functions}
\end{equation}
where $M$ and $\ell$ are integration constants characterizing the EB wormhole. The parameter $M$ is associated with the asymptotic mass of the wormhole, while $\ell$ determines the characteristic length scale of the wormhole and, in particular, its throat size. Following the analysis of~\cite{Daniels:1993yi,Shore:1995fz,Daniels:1995yw,Chen:2016hil}, we introduce an orthonormal vierbein $e^{a}_{\ \mu}$ defined by $g_{\mu\nu}=\eta_{ab}e^{a}_{\ \mu}e^{b}_{\ \nu}$, together with the antisymmetric combination $U^{ab}_{\ \ \mu\nu}=e^{a}_{\ \mu}e^{b}_{\ \nu}-e^{a}_{\ \nu}e^{b}_{\ \mu}$. The Weyl tensor for the EB wormhole metric can then be expressed in the compact bilinear form
\begin{align}
C_{\mu\nu\rho\sigma}=\mathcal{A}\Big(U^{01}_{\mu\nu}U^{01}_{\rho\sigma}-U^{23}_{\mu\nu}U^{23}_{\rho\sigma}\Bigr)+\mathcal{B}\Big(U^{02}_{\mu\nu}U^{02}_{\rho\sigma}+U^{03}_{\mu\nu}U^{03}_{\rho\sigma}-U^{12}_{\mu\nu}U^{12}_{\rho\sigma}-U^{13}_{\mu\nu}U^{13}_{\rho\sigma}\Big),
\label{eq:weyl-wormhole}
\end{align}
where the six scalar functions $\mathcal{A}$ and $\mathcal{B}$ depend
only on $r$, $M$ and $\ell$; their explicit form is given by
\begin{align}
\label{A}
\mathcal{A} &= -\dfrac{(\ell^2+M(-5M+6r))e^h}{3(\ell^2+r^2)^2},\\
\label{B}
\mathcal{B} &= \dfrac{(2\ell^2-M(M-3r))e^h}{3(\ell^2+r^2)^2}.
\end{align}
In the present EB wormhole spacetime, the mass parameter $M$ characterizes the geometry of the generalized EB wormhole. In particular, the Ellis wormhole is recovered in the limit $M=0$. The terms involving $M$ therefore contribute to the geometric structure of the generalized EB wormhole and affect photon propagation and, consequently, the gravitational lensing properties of the spacetime.

\subsection{Polarization decomposition and the light-cone condition}

To organize the photon equation of motion~\eqref{eq:eom-photon} in the
EB wormhole background, one introduces the momentum-projected combinations \cite{Daniels:1993yi,Shore:1995fz,Daniels:1995yw}
\begin{equation}
l_{\nu}=k^{\mu}U^{01}_{\ \ \mu\nu},\qquad
n_{\nu}=k^{\mu}U^{02}_{\ \ \mu\nu},\qquad
m_{\nu}=k^{\mu}U^{23}_{\ \ \mu\nu},
\label{eq:lnr}
\end{equation}
together with the dependent combinations $p_{\nu}=k^{\mu}U^{12}_{\ \ \mu\nu}$,
$r_{\nu}=k^{\mu}U^{03}_{\ \ \mu\nu}$ and $q_{\nu}=k^{\mu}U^{13}_{\ \ \mu\nu}$,
all orthogonal to $k^{\nu}$. Contracting
Eq.~\eqref{eq:eom-photon} with $l^{\nu}$, $n^{\nu}$ and $r^{\nu}$
reduces the photon equation of motion to a homogeneous linear system for
the three independent polarization amplitudes $a\!\cdot\! l$, $a\!\cdot\! n$
and  $a\!\cdot\! m$,
\begin{equation}
\begin{pmatrix} K_{11} & 0 & 0\\ K_{21} & K_{22} & K_{23}\\ 0 & 0 & K_{33}\end{pmatrix}
\begin{pmatrix} a\cdot l\\ a\cdot n\\ a\cdot m\end{pmatrix}=0 ,
\label{eq:Kmatrix}
\end{equation}
whose coefficients $K_{ij}$ depend on $\mathcal{A}$ and $\mathcal{B}$,
the vierbein components and $k_{\mu}$; their lengthy explicit form is
given by 
\begin{align}
K_{11}={}&
(1+8\gamma\mathcal{A})(g_{00}k^0k^0+g_{11}k^1k^1)+(1+8\gamma\mathcal{B})(g_{22}k^2k^2+g_{33}k^3k^3)-2\beta R_{\mu\nu}k^{\mu}k^{\nu},\\
K_{21}={}&8\gamma(\mathcal{A}-\mathcal{B})\sqrt{g_{11}g_{22}}k^1k^2,\\
K_{22}={}&(1+8\gamma\mathcal{B})(g_{00}k^0k^0+g_{11}k^1k^1+g_{22}k^2k^2+g_{33}k^3k^3)-2\beta R_{\mu\nu}k^{\mu}k^{\nu},
\\
K_{23}={}&-8\gamma(\mathcal{A}-\mathcal{B})\sqrt{-g_{00}g_{33}}k^0k^3,\\
K_{33}={}&(1+8\gamma\mathcal{B})(g_{00}k^0k^0+g_{11}k^1k^1)+(1+8\gamma\mathcal{A})(g_{22}k^2k^2+g_{33}k^3k^3)-2\beta R_{\mu\nu}k^{\mu}k^{\nu}.
\end{align}
The determinant condition (\ref{eq:Kmatrix}) for a non-trivial solution factorizes,
\begin{align}
\det K=\tilde K_{11}\,\tilde K_{22}\,\tilde K_{33}=0 .
\label{eq:Kdet-eq}
\end{align}
The first root, $\tilde K_{11}=0$, namely
\begin{align}
(1+8\gamma\mathcal{A})g_{00}k^0k^0+[(1+8\gamma\mathcal{A})g_{11}-2\beta R_{11}]k^1k^1+(1+8\gamma\mathcal{B})(g_{22}k^2k^2+g_{33}k^3k^3)=0.
\label{eq:PPL}
\end{align}
This mode corresponds to a photon whose polarization vector lies within the orbital plane. Following Refs.~\cite{Chen:2015cpa,Lu:2016gsf}, we refer to this polarization as the \emph{PPL} mode. The second root, $\tilde K_{22}=0$, simply reproduces the uncoupled light
cone $\gamma_{\mu\nu}k^{\mu}k^{\nu}=0$ and corresponds to an unphysical polarization
that is discarded, as discussed in~Refs \cite{Drummond:1979pp,Chen:2015cpa,Lu:2016gsf}. 
The third root, $\tilde K_{33}=0$,
\begin{align}
(1+8\gamma\mathcal{B})g_{00}k^0k^0+[(1+8\gamma\mathcal{B})g_{11}-2\beta R_{11}]k^1k^1+(1+8\gamma\mathcal{A})(g_{22}k^2k^2+g_{33}k^3k^3)=0,
\label{eq:PPM}
\end{align}
corresponds to a polarization vector is (effectively) aligned with $l_{\mu}$, i.e.\ orthogonal to the orbital plane; following Ref.~\cite{Chen:2016hil} we denote this mode \emph{PPM}. following the notation of Refs.~\cite{Chen:2015cpa,Lu:2016gsf}.

Equations~\eqref{eq:PPL} and \eqref{eq:PPM} are the two physical light-cone conditions for a curvature-coupled photon propagating in the equatorial plane of an EB wormhole. They define two distinct effective metrics for the two polarization states, so that PPL and PPM photons emitted from the same source generically follow different trajectories, providing a gravitational analogue of birefringence. In the present case, both Weyl and Ricci curvature couplings contribute to the modified photon propagation, leading to a richer structure of the effective light cones than in the case with Weyl coupling alone. In the limit where the curvature couplings $\beta, \gamma$ are switched off, both conditions reduce to the standard light-cone condition of the underlying EB wormhole geometry, and the distinction between the two polarization states disappears. These two light-cone conditions provide the starting point for constructing the effective optical metrics and analyzing the resulting gravitational lensing of curvature-coupled photons.

The two polarization modes give rise to distinct effective optical metrics. In particular, the PPL and PPM modes are described by different effective metrics, reflecting the polarization dependence of photon propagation induced by the curvature couplings. The corresponding effective metrics are given by
\begin{align}
\label{metric}
ds^2&=-e^{h}dt^2+e^{-h}\left[\left(1-4\beta \dfrac{(\ell^2+M^2)e^{h}}{(\ell^2+r^2)^2}\right)dr^2+(\ell^2+r^2)\left(\dfrac{1+8\gamma\mathcal{A}}{1+8\gamma\mathcal{B}}\right)^s(d\theta^2+\sin^2\theta d\phi^2)\right],
\end{align}
where the functions $h$, $\mathcal{A}$, and $\mathcal{B}$ are given by Eqs. (\ref{eq:EB-functions}), (\ref{A}), and (\ref{B}) respectively, and the parameter $s=\pm1$ characterizes the polarization state, with $s=+1$ and $s=-1$ corresponding to the PPL and PPM modes, respectively. These metric functions incorporate the higher-curvature corrections up to the order considered in the effective field theory expansion. In the following, we use these effective optical metrics to investigate the gravitational lensing of photons in the EFT-corrected EB wormhole spacetime.

\section{Strong-Deflection Limit in EFT-Corrected Ellis Wormhole}
\label{sec:strong lens}

In this section, we investigate the deflection of light in an Ellis wormhole spacetime in the presence of higher-curvature corrections arising from the effective field theory (EFT). In the EFT framework, higher-curvature operators modify the propagation of photons through curvature-dependent corrections to the light-cone condition, thereby altering their trajectories from those of minimally coupled photons. We take these modified propagation laws into account and study the resulting light deflection in the strong-deflection regime. In particular, we derive the corresponding deflection angle and lensing observables perturbatively in the EFT couplings. We then compare these observables with those of the Schwarzschild black hole, with the aim of identifying characteristic signatures that could distinguish an EFT-corrected Ellis wormhole from the standard black-hole scenario through gravitational lensing observations.

We first consider the strong-deflection limit of light propagation in the Ellis wormhole spacetime in the absence of EFT corrections, following the analysis of Ref.~\cite{Clement:1982ej,Nandi:2006ds,Dey:2008kn,Muller:2008zza,Bhattacharya:2010zzb,Gibbons:2011rh,Nakajima:2012pu,Tsukamoto:2012np,Tsukamoto:2016qro,Tsukamoto:2016jzh,Tsukamoto:2017edq,Jusufi:2017gyu,Cai:2023ite}. This provides a reference result for the gravitational lensing observables, which will subsequently be used to assess the effects of higher-curvature corrections. In particular, we derive the deflection angle and the corresponding lensing observables in the strong-deflection regime. We then incorporate the modified photon propagation induced by the higher-curvature operators and investigate how the EFT corrections modify these observables. 

The trajectory of photons is described by
\begin{equation}
k^\mu k_\mu = 0,
\end{equation}
where $k^\mu$ is the photon wave number. Equation (\ref{eq:EB-metric}) can be expressed by
\begin{equation}
\label{impact}
\left( \frac{dr}{d\phi} \right)^2
= \frac{1}{b^2} - \frac{1}{r^2 + \ell^2},
\end{equation}
where $b \equiv L/E$ is the impact parameter and
$E \equiv -g_{\mu\nu} t^\mu k^\nu$ and
$L \equiv g_{\mu\nu} \phi^\mu k^\nu$
are the conserved energy and angular momentum of the photon, respectively.
We assume $L \geq 0$ and then $b \geq 0$. The photon is scattered if $b > \ell$
while it falls into the throat if $b < \ell$. In the limit $b \to b_c \equiv \ell$,
the light ray winds around the throat at $r = 0$. The throat is coincident
with a light sphere. We only consider the scatter case, i.e., $b > \ell$.

From $dr/d\phi = 0$ and Eq.~(\ref{impact}), the closest distance is given by
\begin{align}
r_0 = \sqrt{b^2 - \ell^2}.
\end{align}

From Eq.~(\ref{impact}), we obtain the deflection angle of light as
\begin{align}
\alpha = 2 \int_{r_0}^{\infty}
\frac{b\,dr}{\sqrt{(r^2 + \ell^2)(r^2 + \ell^2 - b^2)}} - \pi.
\end{align}

Introducing $x \equiv \frac{b}{\sqrt{r^2 + \ell^2}}$ and $k \equiv \frac{\ell}{b}$,
the deflection angle can be rewritten as
\begin{align}
\alpha = 2K(k) - \pi,
\end{align}
where $K(k)$ is the complete elliptic integral of the first kind defined by
\begin{align}
K(k) = \int_0^1 \frac{dx}{\sqrt{(1 - x^2)(1 - k^2 x^2)}},
\end{align}
with $0 < k < 1$.

The deflection angle can be expanded as
\begin{align}
\alpha = \pi \sum_{n=1}^{\infty}
\left(\frac{(2n-1)!!}{(2n)!!}\right)^2 k^{2n}.
\end{align}

Under the weak-field approximation $\ell \ll b$, the deflection angle becomes
\begin{align}
\alpha = \frac{\pi}{4}k^2 + \frac{9\pi}{64}k^4 + O(k^6),
\end{align}
and we can rewrite it as
\begin{align}
\alpha = \frac{\pi \ell^2}{4 r_0^2}
- \frac{7\pi \ell^4}{64 r_0^4}
+ O\left( \frac{\ell^6}{r_0^6} \right).
\end{align}

In the strong deflection limit $k \to 1$, $K(k)$ behaves as
\begin{align}
\lim_{k \to 1} K(k)
= -\frac{1}{2}\log(1 - k) + \frac{3}{2}\log 2
+ O\big((1-k)\log(1-k)\big).
\end{align}
Thus, the deflection angle in the strong deflection limit $b \to b_c = a$ is
\begin{align}
\alpha(b)
= -\log\left(\frac{b}{b_c} - 1\right)
+ 3\log 2 - \pi
+ O\big((b - b_c)\log(b - b_c)\big).
\end{align}
Therefore, we obtain
\begin{align}
\bar{a} = 1, \quad \bar{b} = 3\log 2 - \pi.
\end{align}

Having established the gravitational lensing observables in the absence of EFT corrections, we now turn to the effects of curvature-photon couplings on photon propagation and the resulting lensing observables. We consider the Ellis wormhole metric given in Eq.~\eqref{metric} for $M=0$. In the presence of curvature-photon couplings, the modified light-cone conditions indicate that the propagation of photons is generally non-geodesic with respect to the background EB wormhole metric. Nevertheless, their trajectories can be described as null geodesics of a polarization-dependent effective metric $\gamma_{\mu\nu}$, satisfying
\begin{align}
\gamma_{\mu\nu}k^\mu k^\nu=0.
\end{align}

From Eq.~(\ref{impact}), we obtain the deflection angle of light as
\begin{align}
\alpha = 2 \int_{r_0}^{\infty}
\frac{b\,dr}{\sqrt{(r^2 + a^2)(r^2 + a^2 - b^2)}}\left(1-\dfrac{2(\beta-4s\gamma)\ell^2}{ (r^2+\ell^2)^2}\right)- \pi.
\end{align}

Using the variables $x \equiv \frac{b}{\sqrt{r^2 + a^2}}$ and $k \equiv \frac{a}{b}$ introduced earlier, the deflection angle can be rewritten as
\begin{align}
\alpha =&\ 2\int_0^1 \frac{dx}{\sqrt{(1 - x^2)(1 - k^2 x^2)}}\left(1-\dfrac{2(\beta-4s\gamma)k^4 x^4}{\ell^2}\right)-\pi\notag\\
 =&\ 2K(k) -\frac{4(\beta-4s\gamma)}{3\ell^2}\left((2+k^2)K(k)-2(1+k^2)E(k)\right)- \pi,
\end{align}
where $E(k)$ is the complete elliptic integral of the second kind defined by
\begin{align}
E(k)=& \int_0^1 \frac{\sqrt{1-k^2x^2}}{\sqrt{1 - x^2}}dx,
\end{align}
with $0 < k < 1$.

The deflection angle can be expanded as
\begin{align}
\alpha =& \pi \sum_{n=1}^{\infty}
\left(\frac{(2n-1)!!}{(2n)!!}\right)^2 k^{2n}-\frac{4(\beta-4s\gamma)}{3\ell^2}\left((2+k^2)K(k)-2(1+k^2)E(k)\right)\notag\\
=&\pi \sum_{n=1}^{\infty}
\left(\frac{(2n-1)!!}{(2n)!!}\right)^2 k^{2n}-\frac{2(\beta-4s\gamma)}{3\ell^2}\left((2+k^2)\pi\sum\limits_{n=0}^{\infty}\left(\frac{(2n-1)!!}{(2n)!!}\right)^2 k^{2n}-2(1+k^2)\pi\sum\limits_{n=0}^\infty\frac{1}{1-2n}\left(\frac{(2n-1)!!}{(2n)!!}\right)^2 k^{2n}\right).
\end{align}

Under the weak-field approximation $\ell \ll b$, the deflection angle becomes
\begin{align}
\alpha = \frac{\pi}{4}k^2 + \frac{9\pi}{64}k^4 -\frac{2(\beta-4s\gamma)}{8\ell^2}\left(3\pi+\frac{5\pi k^2}{4}+\frac{105\pi k^4}{128}\right) + O(k^6),
\end{align}
and we can rewrite it as
\begin{align}
\alpha = \frac{\pi \ell^2}{4 r_0^2}
- \frac{7\pi \ell^4}{64 r_0^4}
-\frac{2(\beta-4s\gamma)}{8\ell^2}\left(3\pi+\frac{5\pi \ell^2}{4r_0^2}-\frac{55\pi \ell^4}{128r_0^4}\right)+ O\left( \frac{\ell^6}{r_0^6} \right).
\end{align}

To investigate the strong-deflection behavior of the bending angle, we now consider the limit in which the impact parameter approaches its critical value, $b\to b_c$. In this limit, the deflection angle develops a logarithmic divergence associated with the unstable circular photon orbit. Following the standard strong-deflection analysis, we separate the deflection integral into divergent and regular parts and evaluate their respective contributions at the critical impact parameter.
\begin{align}
\lim_{b\rightarrow b_c}\alpha(b)=I_D(b_c)+I_R(b_c),
\end{align}
where
\begin{align}
I_D(b_c)=&\lim_{k\rightarrow1}\dfrac{2}{\sqrt{2(1+k)}}\left(1-\frac{2(\beta-4s\gamma)}{\ell^2}\right)\int_0^1\dfrac{dx}{\sqrt{(1-k)(1-x)+k(1-x)^2}}\notag\\
=&\lim_{k\rightarrow1}\dfrac{2}{\sqrt{2(1+k)}}\left(1-\frac{2(\beta-4s\gamma)}{\ell^2}\right)\left(-\frac{1}{\sqrt{k}}\log(1-k)+\frac{2}{\sqrt{k}}\log(1+\sqrt{k})\right),\\
I_R(b_c)=&2 \int_{r_0}^{\infty}
\frac{b_c\,dr}{\sqrt{(r^2 + a^2)(r^2 + a^2 - b_c^2)}}\left(1-\dfrac{2(\beta-4s\gamma)\ell^2}{ (r^2+\ell^2)^2}\right)-I_D(b_c)-\pi.
\end{align}
Thus, the deflection angle in the strong deflection limit \(b \to b_c = \ell\) is given by
\begin{align}
\lim_{b\rightarrow b_c}\alpha(b) = -\left(1-\frac{2(\beta-4s\gamma)}{\ell^2}\right)\log\left(\frac{b}{b_c} - 1\right) + 3\log 2- \frac{2(\beta-4\gamma)}{\ell^2}\left(3\log2-\frac{8}{3}\right)- \pi + \mathcal{O}\bigl((b - b_c)\log(b - b_c)\bigr).
\end{align}
Therefore, we obtain
\begin{align}
\bar{a} = 1-\frac{2(\beta-4s\gamma)}{\ell^2}, \qquad \bar{b} =-\frac{2(\beta-4s\gamma)}{\ell^2}\left(3\log2-\frac{8}{3}\right)- \pi .
\end{align}

In this section, we first reviewed the analysis of the deflection angle for photon propagation in the Ellis wormhole spacetime. We then extended the analysis by incorporating EFT corrections arising from nonminimal couplings between the electromagnetic field and the spacetime curvature. Since EFT corrections are generally expected to be small, it may be difficult to detect them directly through observations of weak gravitational lensing. In the strong-deflection regime, however, the deflection angle exhibits a logarithmic divergence as the impact parameter approaches its critical value, potentially enhancing the sensitivity to such small corrections. More importantly, the curvature-photon coupling $R_{\mu\nu}F^{\mu\rho}F^{\nu}_{\ \rho}$ vanishes in the Schwarzschild spacetime, which is a vacuum solution satisfying $R_{\mu\nu}=0$, while it can contribute in the EB wormhole due to its nonvanishing Ricci curvature. This difference may leave a characteristic imprint on strong-lensing observables, suggesting that strong gravitational lensing could provide a useful means of distinguishing an Ellis wormhole from a Schwarzschild black hole through EFT-induced modifications of photon propagation, even when the EFT corrections are parametrically small.

Building on these results, in the next chapter we extend our analysis to a more general Ellis–Bronnikov wormhole spacetime, with particular emphasis on the strong-deflection limit, where enhanced sensitivity to EFT corrections is expected. By studying the strong-deflection angle in the generalized EB wormhole in detail, we clarify how the curvature-photon coupling and the resulting modification of the photon propagation law affect gravitational lensing observables. In particular, we investigate whether the characteristic EFT corrections associated with the nonvanishing Ricci curvature of the wormhole can provide an observable signature that distinguishes the generalized EB wormhole from the Schwarzschild black hole.

\section{Strong-Deflection Limit in EFT-Corrected Ellis-Bronbikov Wormhole}
\label{sec:strong lens}

In the previous section, we studied the strong-deflection limit of light propagation in the Ellis wormhole spacetime with EFT corrections, providing the baseline lensing observables. In this section, we extend the analysis of the EFT-corrected Ellis wormhole developed in the previous section to the Ellis-Bronnikov (EB) wormhole spacetime. These EFT corrections modify the photon propagation law and, consequently, the effective geometry experienced by photons. We investigate how these modifications affect the strong-deflection lensing observables and assess their potential to distinguish the EB wormhole from the Schwarzschild black hole.

We consider a general static and spherically symmetric spacetime and restrict the photon motion to the equatorial plane, $\theta=\pi/2$. The metric on the equatorial plane can then be written as
\begin{align}
ds^2 = -A(r) dt^2 + B(r) dr^2 + C(r) d\phi^2.
\end{align}

\subsection{Divergent term of the deflection angle}

In this subsection, we review the strong deflection limit analysis of gravitational lensing developed by Valerio Bozza \cite{Bozza:2002zj}. We define new variable
\begin{align}
z = \frac{A(r) - A_0}{1 - A_0},
\end{align}
where $A_0 = A(r_0)$. The orbit equation becomes
\begin{align}
\frac{d\phi}{dr}\equiv I(r_0) = \int_{0}^{1} R(z,r_0)\, f(z,r_0)\, dz,
\end{align}
with
\begin{align}
R(z,r_0) &= \frac{2 \sqrt{AB}}{CA'} (1 - A_0)\sqrt{C_0}, \\
f(z,r_0) &= \frac{1}{\sqrt{A_0-A\frac{C_0}{C}}}.
\end{align}
Here, all functions with the subscript $0$ are evaluated at $r_0$ and ${}^\prime$ denotes differentiation with respect to $r$.

The function $R(z, r_0)$ is regular for all values of $z$ and $r_0$, whereas $f(z, r_0)$ diverges in the limit $z \to 0$. To determine the leading behavior of this divergence in the integrand, we expand the argument of the square root in $f(z, r_0)$ up to second order in $z$:
\begin{align}
\label{pole expansion}
f(z,r_0) \sim f_0(z,r_0) = \frac{1}{\sqrt{p(r_0) z + q(r_0) z^2}}.
\end{align}

When $p(r_0)$ remains nonvanishing, the leading divergence of $f_0$ scales as $z^{-1/2}$, which is integrable and therefore leads to a finite contribution. In contrast, if $p(r_0)$ vanishes, the divergence instead scales as $z^{-1}$, causing the integral to diverge.

From the structure of $p(r_0)$, one finds that it becomes zero at $r_0 = r_{\mathrm{ph}}$, where $r_{\mathrm{ph}}$ denotes the radius of the photon sphere. The photon sphere is defined by the condition
\begin{align}
\label{ph radius}
\frac{d}{dr}\left(\dfrac{C(r)}{A(r)}\right)\bigg|_{r=r_{\rm ph}}= 0.
\end{align}
as discussed in Refs.~\cite{Claudel:2000yi,Teo:2003ltt}.

The photon sphere is located at the radius $r=r_{\mathrm{ph}}$, where photons can follow unstable circular orbits. At the corresponding critical impact parameter $u_c$, the deflection angle becomes divergent in the strong-deflection limit. For photon trajectories with $r_0<r_{\mathrm{ph}}$, the photons are captured by the central region of the wormhole and do not return to the asymptotic region.

To evaluate the integral, we decompose it into two contributions:
\begin{align}
I(r_0) = I_D(r_0) + I_R(r_0),
\end{align}
where
\begin{align}
I_D(r_0) &= \int_{0}^{1} R(0,r_{ph})\, f_0(z,r_0)\, dz,
\end{align}
captures the divergent behavior, and
\begin{align}
\label{regular angle}
I_R(r_0) &= \int_{0}^{1} g(z,r_0)\, dz,
\end{align}
with
\begin{align}
g(z,r_0) = R(z,r_0) f(z,r_0) - R(0,r_{\rm ph}) f_0(z,r_0),
\end{align}
represents the original integrand with the singular contribution removed.

We proceed by computing each part independently and then combining the results to reconstruct the deflection angle. In this subsection, we focus on $I_D$ and its divergence, while in the following subsection we now examine how the finite contribution is obtained.

The integral $I_D(r_0)$ admits an exact evaluation:
\begin{align}
I_D(r_0) = \frac{2R(0,r_{\rm ph})}{\sqrt{q(r_0)}} 
\log \left(\frac{\sqrt{q(r_0)} + \sqrt{p(r_0) + q(r_0)}}{\sqrt{p(r_0)}}\right).
\end{align}

Here, all functions with the subscript ${}_{\rm ph}$ are evaluated at $r_{\rm ph}$.

Substituting these expressions into $I_D(r_0)$ and reorganizing terms, we obtain
\begin{align}
I_D(r_0)= - a \log\left( \frac{r_0}{r_{\rm ph}} - 1 \right)+ b_D+ \mathcal{O}((r_0 - r_{\rm ph})\log(r_0-r_{\rm ph})),
\end{align}
with
\begin{align}
a &= \frac{R(0,r_{\rm ph})}{\sqrt{q(r_{\rm ph})}}, \\
b_D &= \frac{R(0,r_{\rm ph})}{\sqrt{q(r_{\rm ph})}}
\log \left( \frac{2(1 - A_{\rm ph})}{A'_{\rm ph} r_{\rm ph}} \right).
\end{align}

This result provides the leading divergent contribution to the deflection angle, which is logarithmic in nature, as expected.

\subsection{Regular contribution to the deflection angle}

To determine the full coefficient $b$ relevant for strong-field gravitational lensing, following the method developed by Bozza \cite{Bozza:2002zj}, we must supplement $b_D$ with the corresponding contribution from the regular part of the integral defined in Eq.~(\ref{regular angle}).

We expand $I_R(r_0)$ in powers of $(r_0 - r_{ph})$:
\begin{align}
I_R(r_0)
=
\sum_{n=0}^{\infty}
\frac{1}{n!} (r_0 - r_{\rm ph})^n
\int_{0}^{1}
\left.
\frac{\partial^n g}{\partial r_0^n}
\right|_{r_0 = r_{\rm ph}}
dz.
\end{align}

Without subtracting the divergent part from $R(z,r_0)f(z,r_0)$, the coefficient corresponding to $n=0$ would diverge, whereas all higher-order terms would be finite. By construction, however, the function $g(z,r_0)$ is regular at $z=0$ and $r_0 = r_{ph}$, as can be verified through a series expansion, using $p(r_{\rm ph}) = 0$.

Restricting ourselves to terms up to $\mathcal{O}(r_0 - r_{\rm ph})$, we keep only the leading term:
\begin{align}
I_R(r_0)
=
\int_{0}^{1} g(z,r_{\rm ph})\, dz
+ \mathcal{O}(r_0 - r_{\rm ph}),
\end{align}
and define
\begin{align}
b_R = I_R(r_{\rm ph}).
\end{align}

Including also the $-\pi$ contribution in the deflection angle, we arrive at
\begin{align}
b = -\pi + b_D + b_R.
\end{align}

The quantity $b_R$ can be evaluated numerically for general metrics, since the integrand is regular. In several cases, however, an analytic expression can be obtained. For instance, in the Schwarzschild spacetime, the integral can be computed exactly. More broadly, one may expand the integral around the Schwarzschild limit in terms of metric parameters, allowing each term in the expansion to be determined separately.

In the general formula for the deflection angle, the impact parameter $u$ is introduced. We begin by defining it as a function of the closest approach distance $r_0$:
\begin{align}
\label{impact para}
u = \sqrt{\dfrac{C_0}{A_{0}}}.
\end{align}

From Eq.~(\ref{ph radius}), the minimum impact parameter is given by
\begin{align}
u_{\rm ph} = \sqrt{\dfrac{C_{\rm ph}}{A_{\rm ph}}}.
\end{align}

Expanding Eq.~(\ref{impact para}) yields
\begin{align}
u - u_{\rm ph}= c (r_0 - r_{\rm ph})^2+\mathcal{O}\left( (r_0 - r_{\rm ph})^3\right).
\end{align}

Using this relation, the deflection angle may be expressed as a function of the angular position $\theta$:
\begin{align}
\alpha(\theta)
=
- \bar{a} \log\left( \frac{\theta D_{OL}}{u_{\rm ph}} - 1 \right)
+ \bar{b},
\end{align}
with
\begin{align}
\bar{a} &= \frac{a}{2} = \frac{R(0,r_{\rm ph})}{2\sqrt{q(r_{\rm ph})}}, \\
\bar{b} &= b + \frac{a}{2} \log \left( \frac{c r_{\rm ph}^2}{u_{\rm ph}} \right)
= -\pi + b_R + \bar{a} \log \left( \frac{2q(r_{\rm ph})}{A_{\rm ph}} \right).
\end{align}

\subsection{Strong Deflection Analysis}

In this subsection, we analyze gravitational lensing in the strong-deflection regime, taking into account the effects of the curvature-photon couplings introduced in Eq.~\eqref{eq:action}. We consider the Ellis-Bronnikov (EB) wormhole and set the throat-size parameter to $\ell=1$. This choice does not restrict the generality of our analysis, since $\ell$ sets the characteristic length scale of the wormhole and can be absorbed by an appropriate rescaling of the coordinates and other dimensional quantities. We therefore keep the mass parameter $M$ arbitrary throughout the analytical derivation.

For photons propagating in the equatorial plane, the EFT-corrected propagation law can be recast in terms of an effective metric. In the present case, the effective metric takes the form
\begin{align}
ds^2&=-e^{h}dt^2+e^{-h}\left[\left(1-4\beta \dfrac{(1+M^2)e^{h}}{(1+r^2)^2}\right)dr^2+(1+r^2)\left(\dfrac{1+8\gamma\mathcal{A}}{1+8\gamma\mathcal{B}}\right)^s(d\theta^2+\sin^2\theta d\phi^2)\right],\\
h(r)&=2M\left[\arctan\left(r\right)-\frac{\pi}{2}\right],\\
\mathcal{A} &= -\dfrac{e^{-M(\pi-2\arctan r)}(1+M(-5M+6r))}{3(1+r^2)^2},\\
\mathcal{B} &= \dfrac{e^{-M(\pi-2\arctan r)}(2-M(M-3r))}{3(1+r^2)^2}.
\end{align}
The strong-deflection analysis then allows us to derive the lensing observables while retaining the dependence on the physical parameters. In particular, the deflection angle is decomposed into a logarithmically divergent part and a regular contribution. The divergent part can be obtained analytically, whereas the regular contribution $b_R$ involves an integral that cannot be evaluated in closed form for the present setup. We therefore evaluate $b_R$ numerically for several representative parameter choices.

For the spacetime under consideration, the radius of the photon sphere is given by
\begin{equation}
r_{\rm ph}=2M-\dfrac{12s\gamma Me^{-M(\pi-2\arctan 2M)}}{(1+4M^2)}.
\end{equation}
The photon-sphere radius plays a central role in determining the properties of strong gravitational lensing. In particular, it determines the critical impact parameter $b_c$, which characterizes the boundary between photons that are scattered back to the observer and those that are captured by the central object. The critical impact parameter is directly related to the angular position of the relativistic images and therefore provides a natural connection between the photon-sphere structure and observable lensing quantities. We thus use the photon-sphere radius obtained above as a key ingredient in the subsequent analysis of the strong-deflection limit.

In this background, we evaluate the deflection angle in the strong deflection limit following the standard formalism. The relevant functions $R(z,r_{ph})$ and $f(z,r_{ph})$ appearing in the integral expression of the deflection angle are given by
\begin{align}
R(z,r_{ph})=&\frac{\left(1-e^{-M(\pi-2\arctan(2M))}\right)\sqrt{e^{M(\pi-2\arctan(2M))}(1+4M^2)}}{M}\notag\\
&\quad-\frac{2\beta}{M}\frac{e^{-M\left[\pi+2\arctan(\cot\Theta)\right]}\left(1-e^{-M(\pi-2\arctan(2M))}\right)(1+M^2)\sqrt{e^{M(\pi-2\arctan(2M))}(1+4M^2)}}{\left(1+\cot^2\Theta\right)^2}\notag\\
&\quad-\frac{4s\gamma}
{\sqrt{e^{M(\pi-2\arctan(2M))}(1+4M^2)}\,(M+4M^3)}
e^{-2M\left[\pi-\arctan(2M)+\arctan(\cot\Theta)\right]}
\notag\\
&\quad\times
\Biggl\{
e^{M\left[\pi-2\arctan(2M)+2\arctan(\cot\Theta)\right]}
\Bigl[
e^{M\pi}(1+7M^2)
-e^{2M\arctan(2M)}(1+13M^2)
\Bigr]
\notag\\
&\qquad\quad
+2e^{M(\pi-2\arctan(2M))}
\left[-1+e^{M(\pi-2\arctan(2M))}\right]
(-1+2M^2)(1+4M^2)^2
\sin^4\Theta
\notag\\
&\qquad\quad
+3e^{M(\pi-2\arctan(2M))}
\left[-1+e^{M(\pi-2\arctan(2M))}\right]
M(1+4M^2)^2
\sin^2\Theta\,
\sin(2\Theta)
\Biggr\},\\
f(z,r_{\rm ph})=&\Biggl[e^{-M[\pi-2\arctan(2M)]}-(1+4M^2)e^{-M[\pi-2\arctan(2M)]-2M\arctan(\cot\Theta)}\sin^2\Theta\left\{1+
\left(e^{M[\pi-2\arctan(2M)]}-1\right)z\right\}\Biggr]^{-1/2}\notag\\
&-\frac{2s\gamma}{(1+4M^2)^2}
\Biggl[
e^{-2M\left[\pi-\arctan(2M)+\arctan(\cot\Theta)\right]}
\notag\\
&\quad\times
\Biggl\{
e^{M\left[\pi+2\arctan(\cot\Theta)\right]}
-e^{M(\pi-2\arctan(2M))}
(1+4M^2)\sin^2\Theta
\left[
1+\left(e^{M(\pi-2\arctan(2M))}-1\right)z
\right]
\Biggr\}
\Biggr]^{-1/2}
\notag\\
&\quad\times
\frac{1}{
e^{-M(\pi-2\arctan(2M))}
-e^{-M\left[\pi+2\arctan(\cot\Theta)\right]}
(1+4M^2)\sin^2\Theta
\left\{1+\left[e^{M(\pi-2\arctan(2M))}-1\right]z\right\}
}
\notag\\
&\quad\times
\Biggl\{
-6e^{-2M(\pi-2\arctan(2M))}M^2
\nonumber\\
&\qquad
-6e^{-2M\left[\pi-\arctan(2M)+\arctan(\cot\Theta)\right]}
M^2(1+4M^2)\sin^2\Theta\,(z-1)
\notag\\
&\qquad
+e^{-M\left[3\pi-2\arctan(2M)+4\arctan(\cot\Theta)\right]}
(1+4M^2)
\Biggl[
-3e^{M\left[\pi+2\arctan(\cot\Theta)\right]}
M\sin(2\Theta)(z-1)
\notag\\
&\qquad
+2e^{M(\pi-2\arctan(2M))}
(-1+2M^2)(1+4M^2)^2
\sin^6\Theta
\left\{1+\left[e^{M(\pi-2\arctan(2M))}-1\right]z\right\}
\notag\\
&\qquad
+3e^{M(\pi-2\arctan(2M))}
M(1+4M^2)^2
\sin^4\Theta\,\sin(2\Theta)
\left\{1+\left[e^{M(\pi-2\arctan(2M))}-1\right]z\right\}
\notag\\
&\qquad
+2e^{M\left[\pi-2\arctan(2M)+2\arctan(\cot\Theta)\right]}
\sin^2\Theta
\Bigl[
-e^{2M\arctan(2M)}(1+10M^2)(z-1)
+e^{M\pi}(1+7M^2)z
\Bigr]
\Biggr]
\Biggr\},\\
\Theta(z) \equiv&\frac{1}{2M}\log\!\left[-e^{-M(\pi-2\arctan(2M))}(z-1)+z\right],
\end{align}
where $z$ is defined in terms of the radial coordinate $R$.

Substituting these expressions into the deflection angle formula and performing the integration, we obtain the regular part of the deflection angle as
\begin{table}[H]
\centering
\caption{Numerical values of the $\beta$- and $\gamma$-dependent contributions to the regular part $b_R$ for representative values of the mass parameter $M$.}
\label{tab:bR}
\begin{tabular}{c|c|c|c}
\hline
$M$ & $b_R^{(0)}$ & $b_R^{(\beta)}$ & $b_R^{(\gamma)}$ \\
\hline
 0.3 & 0.722358 & 2.28141 & -0.669665\\
\hline
 0.5 & 0.79729 & 1.13164 & -0.733851\\
\hline
 1 & 0.864372 & 0.271682 & -0.506986\\
\hline
 1.5 & 0.882856 & 0.108379 & -0.29205\\
\hline
 2 & 0.890058 & 0.0572838 & -0.181255\\
\hline
 5 & 0.898337 & 0.0083697 & -0.0324116\\
\hline
\end{tabular}
\end{table}

We next determine the strong deflection limit coefficients and the photon sphere quantities. The corresponding expressions are given by
\begin{align}
q_{\rm ph}&=\dfrac{(1+4M^2)(1-e^{M(\pi-\arctan 2M)})^2e^{-M(\pi-\arctan 2M)}}{4M^2}\notag\\
&\qquad+\dfrac{2s\gamma(2-7M^2-(2-M^2)e^{M(\pi-2\arctan 2M)})(1-e^{M(\pi-\arctan 2M)})e^{-2M(\pi-\arctan 2M)}}{M^2(1+4M^2)},\\
\bar{a}\ &=\frac{2\left(1-e^{-M(\pi-2\arctan(2M))}\right)\sqrt{e^{M(\pi-2\arctan(2M))}(1+4M^2)}}{\sqrt{e^{-M(\pi-2\arctan(2M))}\left(-1+e^{M(\pi-2\arctan(2M))}\right)^2(1+4M^2)}} \notag \\ 
&-\frac{4\beta\,e^{-2M(\pi-2\arctan(2M))}\left(-1+e^{M(\pi-2\arctan(2M))}\right)(1+M^2)\sqrt{e^{M(\pi-2\arctan(2M))}(1+4M^2)}}
{(1+4M^2)^2}\notag\\
&\quad\times
\Biggl[\frac{e^{-2M(\pi-2\arctan(2M))}\left(-1+e^{M(\pi-2\arctan(2M))}\right)}{1+4M^2}\notag\\
&\quad\times\Biggl\{e^{2M(\pi-2\arctan(2M))}(1+4M^2)^2-e^{M(\pi-2\arctan(2M))}(1+8M^2+16M^4)\Biggr\}\Biggr]^{-1/2}\notag\\
&+\frac{8s\gamma\left(-1+e^{M(\pi-2\arctan(2M))}\right)(-1+2M^2)}{M\left(e^{M(\pi-2\arctan(2M))}(1+4M^2)\right)^{3/2}\sqrt{\dfrac{e^{-M(\pi-2\arctan(2M))}\left(-1+e^{M(\pi-2\arctan(2M))}\right)^2(1+4M^2)}{M^2}}},\\
b_D&=\frac{2\left(-1+e^{M(\pi-2\arctan(2M))}\right)(1+4M^2)\log\!\left(\dfrac{\left(-1+e^{M(\pi-2\arctan(2M))}\right)(1+4M^2)}{2M^2}\right)}{\sqrt{e^{M(\pi-2\arctan(2M))}(1+4M^2)}\sqrt{e^{-M(\pi-2\arctan(2M))}\left(-1+e^{M(\pi-2\arctan(2M))}\right)^2(1+4M^2)}} \notag \\ 
&\quad-\frac{4\beta\,e^{-2M(\pi-2\arctan(2M))}\left(-1+e^{M(\pi-2\arctan(2M))}\right)(1+M^2)\sqrt{e^{M(\pi-2\arctan(2M))}(1+4M^2)}}{(1+4M^2)^2}\notag\\
&\quad\times\log\!\left[\frac{e^{M(\pi-2\arctan(2M))}\left(1-e^{-M(\pi-2\arctan(2M))}\right)(1+4M^2)}{2M^2}\right]\notag\\
&\quad\times\Biggl[\frac{e^{-2M(\pi-2\arctan(2M))}\left(-1+e^{M(\pi-2\arctan(2M))}\right)}{1+4M^2}\notag\\
&\qquad\times\left\{e^{2M(\pi-2\arctan(2M))}(1+4M^2)^2-e^{M(\pi-2\arctan(2M))}(1+8M^2+16M^4)\right\}\Biggr]^{-1/2}\notag\\
&+\frac{4s\gamma\Biggl[3\left(-1+e^{M(\pi-2\arctan(2M))}+4M^2\right)+2\left(-1+e^{M(\pi-2\arctan(2M))}\right)(-1+2M^2)\log\!\left(
\frac{\left(-1+e^{M(\pi-2\arctan(2M))}\right)(1+4M^2)}{2M^2}\right)\Biggr]}{\left[e^{M(\pi-2\arctan(2M))}(1+4M^2)\right]^{3/2}
\sqrt{e^{-M(\pi-2\arctan(2M))}\left(-1+e^{M(\pi-2\arctan(2M))}\right)^2(1+4M^2)}},\\
u_{\rm ph}&=\sqrt{(1+4M^2)e^{2M(\pi-2\arctan 2M)}}-\dfrac{4\gamma e^{M(\pi-2\arctan 2M)}}{\sqrt{(1+4M^2)e^{2M(\pi-2\arctan 2M)}}}.
\end{align}

Combining the divergent and regular contributions, we obtain the coefficient $\bar{b}$ appearing in the deflection angle as
\begin{align}
\bar{b}=-\pi+b_D+b_R+\bar{a}\Biggl(\log\dfrac{2M^2}{1+4M^2}+\dfrac{4s\gamma e^{-M(\pi-2\arctan(2M))}}{1+4M^2}\Biggr)
\end{align}

Finally, the deflection angle in the strong deflection limit can be written in terms of the angular position $\theta$ as
\begin{align}
\alpha(\theta)=&-\Biggl(\frac{2\left(1-e^{-M(\pi-2\arctan(2M))}\right)\sqrt{e^{M(\pi-2\arctan(2M))}(1+4M^2)}}{\sqrt{e^{-M(\pi-2\arctan(2M))}\left(-1+e^{M(\pi-2\arctan(2M))}\right)^2(1+4M^2)}} \notag \\ 
&\quad-\frac{4\beta\,e^{-2M(\pi-2\arctan(2M))}\left(-1+e^{M(\pi-2\arctan(2M))}\right)(1+M^2)\sqrt{e^{M(\pi-2\arctan(2M))}(1+4M^2)}}
{(1+4M^2)^2}\notag\\
&\quad\times
\Biggl[\frac{e^{-2M(\pi-2\arctan(2M))}\left(-1+e^{M(\pi-2\arctan(2M))}\right)}{1+4M^2}\notag\\
&\quad\times\Biggl\{e^{2M(\pi-2\arctan(2M))}(1+4M^2)^2-e^{M(\pi-2\arctan(2M))}(1+8M^2+16M^4)\Biggr\}\Biggr]^{-1/2}\Biggr)\notag\\
&+\frac{8s\gamma\left(-1+e^{M(\pi-2\arctan(2M))}\right)(-1+2M^2)}{M\left(e^{M(\pi-2\arctan(2M))}(1+4M^2)\right)^{3/2}\sqrt{\dfrac{e^{-M(\pi-2\arctan(2M))}\left(-1+e^{M(\pi-2\arctan(2M))}\right)^2(1+4M^2)}{M^2}}}\notag\\
&\quad\times\log\left(\frac{\theta D_{OL}}{\sqrt{(1+4M^2)e^{2M(\pi-2\arctan 2M)}}-\dfrac{4\gamma e^{M(\pi-2\arctan 2M)}}{\sqrt{(1+4M^2)e^{2M(\pi-2\arctan 2M)}}}}-1\right)+\bar{b}.
\end{align}

\begin{figure}[t]
    \centering
    \includegraphics[width=0.5\textwidth]{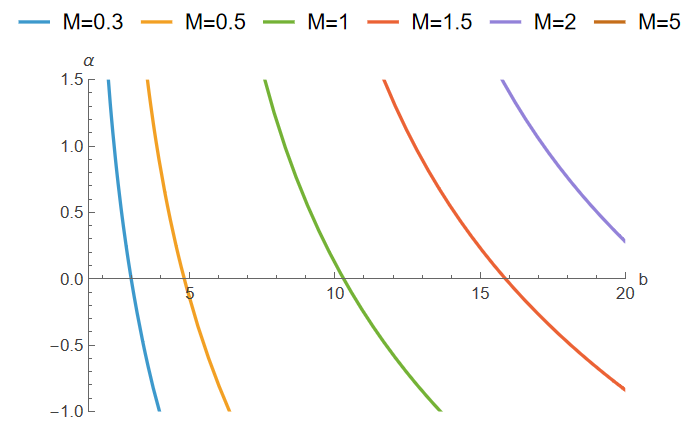}
    \caption{Deflection angle $\alpha(b)$ for the higher-curvature-corrected EB wormhole with $s=1$ (PPL mode), and higher-curvature coupling constants of order $\mathcal{O}(10^{-3})$. For clarity, we plot the deflection angle with the constant term $\bar{b}$ subtracted.}
    \label{fig:EB_deflection_angle}
\end{figure}
\begin{figure}[t]
    \centering
    \includegraphics[width=0.5\textwidth]{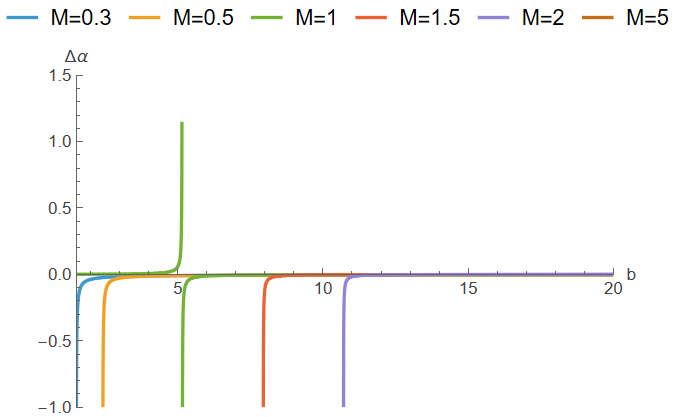}
    \caption{Difference in the deflection angle $\Delta\alpha$ between the higher-curvature-corrected EB wormhole and the standard EB wormhole for $s=1$ (PPL mode), with higher-curvature coupling constants of order $\mathcal{O}(10^{-3})$. For clarity, we plot the deflection angle with the constant term $\bar{b}$ subtracted.}
    \label{fig:EB_delta_deflection_angle}
\end{figure}

The strong-deflection limit corresponds to the case in which the closest approach $r_0$ approaches the radius of the photon sphere $r_{\rm ph}$. In this limit, the deflection angle exhibits a logarithmic divergence, as demonstrated for the Ellis wormhole in the previous chapter, and can be systematically analyzed using the formalism developed in the previous subsection. Since the EFT corrections are treated perturbatively, their contributions to gravitational lensing are generally small and may be difficult to detect in the weak-deflection regime. In the strong-deflection regime, however, the logarithmic behavior can enhance the contribution of these corrections to the deflection angle near the critical impact parameter. This enhancement is illustrated in Fig.~\ref{fig:EB_delta_deflection_angle}, where the difference in the deflection angle $\Delta\alpha$ between the higher-curvature-corrected EB wormhole and the standard EB wormhole becomes significantly enhanced near the critical impact parameter for the PPL mode ($s=1$). The PPM mode ($s=-1$) is expected to exhibit a similar enhancement, since the contribution from a particular EFT correction term differs from that in the PPL mode only by its sign. Therefore, the strong-deflection regime may provide a promising avenue for distinguishing the EB wormhole from black holes through gravitational lensing observations. The logarithmic enhancement can magnify differences in their deflection angles even when the EFT corrections themselves are perturbatively small.

One of the main goals of this work is to identify observational signatures that can distinguish black holes from wormholes through gravitational lensing. In particular, as in the Ellis wormhole analyzed in the previous chapter, the EB wormhole considered here admits a nonvanishing contribution from the curvature-electromagnetic interaction $R_{\mu\nu}F^{\mu\rho}F^\nu_{\ \rho}$, in contrast to the Schwarzschild spacetime, which is a vacuum solution satisfying $R_{\mu\nu}=0$. Although the corresponding EFT correction may be small, its contribution to the logarithmically divergent part of the deflection angle can be enhanced in the strong deflection limit. It may therefore leave a characteristic imprint on the lensing observables, potentially providing an observational distinction between the EB wormhole and the Schwarzschild black hole.

\section{Time Delay in EFT-Corrected Ellis-Bronbikov Wormhole}
\label{sec:time delay}

In this section, we investigate the time delay of photons in the EFT-corrected EB wormhole spacetime. As in the gravitational lensing analysis presented in the previous section, we use the time delay as an observable to explore potential differences between the EB wormhole and the Schwarzschild black hole. In particular, we derive the time delay in the strong-deflection limit and examine how the EFT-induced modifications to photon propagation affect this observable. This analysis allows us to assess whether the time-delay signal can provide an additional observational signature for distinguishing the EB wormhole from a black hole.

\subsection{Travel Time Integral}

The coordinate time along a photon trajectory can be obtained by eliminating the affine parameter from the equations of motion,
\begin{equation}
\frac{dt}{dr} = \frac{\dot{t}}{\dot{r}},
\end{equation}
where $\dot{t}$ and $\dot{r}$ are determined by the conserved energy and angular momentum associated with the static and spherically symmetric spacetime. Using these conserved quantities together with the null condition for photon propagation, the radial derivative of the coordinate time can be written as
\begin{align}
\frac{dt}{dr} =&\dfrac{\sqrt{BA_0}}{\sqrt{A}}\dfrac{1}{\sqrt{A_0-A\frac{C_0}{C}}}.
\end{align}

The travel time along a photon trajectory is then obtained by integrating this expression from the point of closest approach to the observer and source. In analogy with the treatment of the deflection angle, it is useful to reorganize the resulting integral so as to separate the contribution associated with the photon-sphere region from the remaining regular part. This decomposition provides a convenient starting point for analyzing the behavior of the travel time in the strong-deflection regime, where the photon trajectory approaches the unstable photon orbit.

The difference in travel times between two photon trajectories characterized by the closest approach radii $r_{0,1}$ and $r_{0,2}$ can then be expressed as
\begin{align}
 T_1 - T_2 = \tilde{T}(r_{0,1}) - \tilde{T}(r_{0,2}) + 2 \int_{r_{0,1}}^{r_{0,2}} \frac{\sqrt{B(r)}}{\sqrt{A(r)}}\, dr,
\end{align}
where the functions entering this decomposition are defined by
\begin{align}
\tilde{T}(r_0) =&\int_0^1\tilde{R}\dfrac{1}{\sqrt{A_0-A(r(z))\frac{C_0}{C(r(z))}}}dz,\\ \tilde{R}(z,r_{0})= &2\dfrac{1-A_{0}}{A'(r(z))}\dfrac{\sqrt{B(r(z))A_{0}}}{\sqrt{A(r(z))}}\left(1-\dfrac{\sqrt{A_0-A(r(z))\frac{C_0}{C(r(z))}}}{\sqrt{A_0}}\right).
\end{align}
Here, the variable $z$ parametrizes the radial integration domain such that the closest approach is mapped to $z=0$. This form is particularly useful because the divergent behavior associated with the photon sphere is contained in the factor involving $A_0-A(r(z))C_0/C(r(z))$.

As the closest approach radius approaches the photon-sphere radius, the leading contribution to the travel-time integral can be extracted by expanding the relevant functions around $z=0$. The resulting expression takes the canonical form
\begin{equation}
\tilde{T}(u) = -\tilde{a} \log\left(\frac{u}{u_{\rm ph}} - 1\right) + \tilde{b} + \mathcal{O}(u-u_{\rm ph}),
\end{equation}
where $u_{\rm ph}$ denotes the critical impact parameter associated with the photon sphere. The coefficient of the logarithmic term is determined entirely by the behavior of the metric functions in the vicinity of the photon sphere and is given by
\begin{align}
\tilde{a} =\dfrac{\tilde{R}(0,r_{\rm ph})}{2\sqrt{q_{\rm ph}}}.
\end{align}
Thus, the leading term characterizing the logarithmic divergence of the photon travel time in the strong-deflection limit is determined by the local geometrical properties near the photon sphere, while the constant term $\tilde b$ represents the remaining regular contribution.

\subsection{Relation to Deflection Angle}

The strong-deflection expansion of the travel time can be compared directly with that of the deflection angle. In the same limit, the deflection angle takes the form
\begin{equation}
\alpha(u) = -\bar{a} \log\left(\frac{u}{u_{\rm ph}} - 1\right) + \bar{b},
\end{equation}
where the coefficient of the logarithmic divergence is
\begin{equation}
\bar{a} = \frac{R(0,r_{\rm ph})}{2\sqrt{\beta_{\rm ph}}}.
\end{equation}
The analogous logarithmic structures of $\tilde{T}(u)$ and $\alpha(u)$ reflect their common origin in the near-photon-sphere behavior of the photon trajectory. Nevertheless, the corresponding coefficients encode different combinations of the metric functions and therefore provide complementary information about photon propagation. In the EFT-corrected EB wormhole spacetime, both coefficients receive corrections from the modified photon propagation law. Comparing the time delay with the deflection angle therefore provides complementary information that may help identify characteristic differences between the EB wormhole and the Schwarzschild black hole, offering a potential means of distinguishing the two spacetimes through gravitational lensing observables.

\subsection{Time Delay Between Relativistic Images}

In the strong deflection regime, relativistic images correspond to photon trajectories that wind multiple times around the photon sphere of the wormhole. For trajectories with winding number $n$, the corresponding impact parameters satisfy
\begin{equation}
\frac{u}{u_{\rm ph}} - 1 \simeq 
\exp\left(\frac{\bar{b} - 2\pi n}{\bar{a}}\right).
\end{equation}

Substituting this relation into the time expression, we obtain
\begin{equation}
T_n = \tilde{a} \frac{2\pi n - \bar{b}}{\bar{a}} + \tilde{b}.
\end{equation}
Here, $T_n$ denotes the photon travel time for a trajectory that winds $n$ times around the photon sphere.

The time delay between two relativistic images labeled by $n$ and $m$ is therefore given by
\begin{align}
\Delta T_{n,m}\equiv& \ T_n-T_m\notag\\
=&\ 2\pi(n-m)\left|\frac{\tilde{a}}{\bar{a}}\right|+2\sqrt{\dfrac{B_{\rm ph}}{A_{\rm ph}}}\sqrt{\dfrac{u_{\rm ph}}{c}}e^{\frac{\bar{b}}{2\bar{a}}}\left(e^{-\frac{\pi n}{|\bar{a}|}}-e^{\frac{\pi m}{|\bar{a}|}}\right).
\end{align}

Below, we present the analytic results for the time delay in the EFT-corrected EB wormhole spacetime. For simplicity, we set the throat radius to $\ell=1$. The results for both the PPL and PPM modes are given as follows:
\begin{align}
\label{time delay}
\Delta T_{n,m}=&\ 2\pi(n-m)\Biggl(\frac{1}{2}\sqrt{e^{M(\pi-2\arctan(2M))}}\sqrt{e^{M(\pi-2\arctan(2M))}(1+4M^2)}\notag\\
&\times\sqrt{\frac{e^{-M(\pi-2\arctan(2M))}\left(-1+e^{M(\pi-2\arctan(2M))}\right)^2(1+4M^2)}{M^2}}\notag\\ 
&-\frac{2s\gamma
\sqrt{e^{M(\pi-2\arctan(2M))}(1+4M^2)}
\sqrt{\dfrac{e^{-M(\pi-2\arctan(2M))}
\left(-1+e^{M(\pi-2\arctan(2M))}\right)^2(1+4M^2)}
{M^2}}}
{\sqrt{e^{M(\pi-2\arctan(2M))}}
\left(-1+e^{M(\pi-2\arctan(2M))}\right)
(1+4M^2)^2} \notag\\
&\qquad\times
\left(1-11M^2
+e^{M(\pi-2\arctan(2M))}
(-1+5M^2)\right)\Biggr)\notag\\
&+\left(2\sqrt{e^{2M(\pi-2\arctan(2M))}}\sqrt{2+8M^2} -\frac{4\sqrt{2}\,e^{-M(\pi-2\arctan(2M))}\sqrt{e^{2M(\pi-2\arctan(2M))}}(\beta+4s\gamma)(1+M^2)}{(1+4M^2)^{3/2}} \right)\notag\\
&\times e^{\frac{\bar{b}}{2\bar{a}}}\left(e^{-\frac{\pi n}{|\bar{a}|}}-e^{-\frac{\pi m}{\bar{a}}}\right).
\end{align}

\begin{figure}[t]
    \centering
    \includegraphics[width=0.5\textwidth]{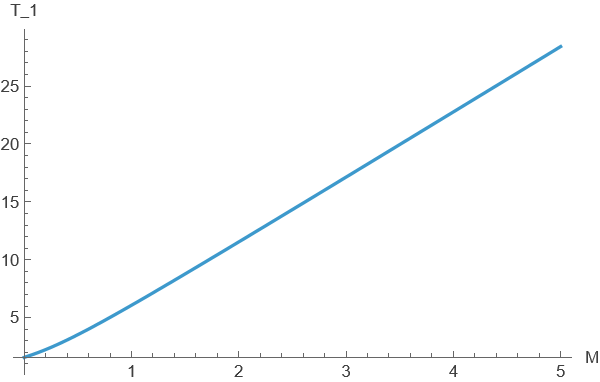}
    \caption{Leading contribution $T_1$ to the time difference $\Delta T_{n,m}$ between two photon paths as a function of $M$ for the EB wormhole, where $T_1$ is the first term in Eq.~(\ref{time delay}).}
    \label{fig:EB_time_delay}
\end{figure}

The leading contribution to the time delay $\Delta T_{n,m}$ is given by the first term $T_1$ in Eq.~(\ref{time delay}), and its dependence on the mass parameter $M$ is shown in Fig.~\ref{fig:EB_time_delay}. The figure indicates that the time delay increases with increasing $M$. These results demonstrate that, for both the PPL and PPM modes, the time delay has the same structural dependence on the winding numbers $(n,m)$ and the strong-deflection coefficients. The distinction between the two polarization modes appears only through the sign of the EFT correction terms, while the leading-order structure of the time delay remains otherwise unchanged.

The time-delay analysis provides a complementary probe of the EFT corrections to photon propagation in the EB wormhole spacetime. Unlike the deflection angle, however, the time delay does not exhibit a logarithmically divergent contribution in the strong-deflection limit. As a result, the characteristic effects of the EFT corrections are less pronounced, making it more challenging to identify clear observational differences between the EB wormhole and the Schwarzschild black hole from the time delay alone. Nevertheless, the time-delay observable provides an independent probe of the modified photon propagation and can complement the gravitational lensing observables discussed in the previous section. A combined analysis of these observables may therefore provide a more comprehensive way to characterize the signatures of EFT-corrected wormhole spacetimes.

\section{Conclusion and Discussion}
\label{sec:conclusion}

In this work, we have investigated photon propagation and gravitational lensing in an Ellis--Bronnikov (EB) wormhole spacetime in the presence of effective field theory (EFT) corrections. Starting from an effective action containing curvature-photon couplings, we derived the modified propagation law for photons and showed that the higher-curvature interactions modify photon trajectories in a polarization-dependent manner. We then recast the modified propagation law in terms of an effective metric, providing a convenient framework for studying photon trajectories and lensing observables in the EFT-corrected EB wormhole spacetime.

We first analyzed the photon trajectories in the strong-deflection limit and derived the corresponding corrections to the photon-sphere radius and the deflection angle. Although the EFT corrections are expected to be parametrically small and may therefore be difficult to detect through gravitational lensing in the weak-field regime, their effects can be enhanced in the strong-deflection limit due to the logarithmic divergence of the deflection angle as the impact parameter approaches its critical value. This enhancement may make the strong-deflection regime particularly sensitive to small EFT corrections. We found that the EFT corrections modify the strong-deflection coefficients and consequently affect the associated lensing observables. In particular, the curvature-photon interaction $R_{\mu\nu}F^{\mu\rho}F^{\nu}_{\ \rho}$ provides a distinctive contribution in the EB wormhole spacetime. This interaction is nonvanishing because the EB wormhole is not Ricci-flat, whereas it vanishes in the Schwarzschild spacetime. Therefore, even though the EFT corrections themselves are small, their enhanced contribution to the logarithmically divergent part of the deflection angle may provide an observational handle for distinguishing the EB wormhole from the Schwarzschild black hole. In this sense, EFT corrections to photon propagation may offer a new way to identify characteristic differences between wormhole and black hole spacetimes through strong gravitational lensing.

The above results indicate that strong-deflection lensing can provide a useful probe of EFT corrections in wormhole spacetimes. Importantly, the EFT corrections are not simply universal shifts of the lensing observables, but depend on the curvature structure of the background spacetime. The polarization dependence of the photon propagation further implies that the curvature-photon couplings can, in principle, induce polarization-dependent lensing signatures. The strong gravitational field near the photon sphere therefore provides a natural regime in which perturbatively small higher-curvature effects may become relevant.

We have also investigated the photon time delay in the strong-deflection regime. Unlike the deflection angle, the time delay does not exhibit a logarithmically divergent contribution of the same form near the photon sphere. As a consequence, the EFT-induced corrections to the time delay are less pronounced, and the time delay alone may not provide a particularly sensitive observable for distinguishing the EB wormhole from the Schwarzschild black hole. Nevertheless, it constitutes an independent probe of the modified photon propagation and provides complementary information to the lensing observables. A combined analysis of the deflection angle, other lensing observables, and time delay may therefore provide a more complete characterization of EFT-corrected wormhole spacetimes.

Several directions remain for future investigation. First, it would be interesting to investigate the observational consequences of the polarization dependence predicted here, particularly its impact on the deflection angle and the resulting polarization-dependent lensing observables. Second, a more realistic treatment of lensing observations would require finite-distance effects, higher-order corrections in the strong-deflection expansion, and a systematic exploration of the allowed EFT parameter space. It would also be interesting to extend the present analysis to rotating wormhole geometries \cite{Jusufi:2017mav,Ono:2018ybw}, where the interplay between curvature-photon couplings and frame dragging may lead to additional characteristic signatures. Such extensions may help clarify the extent to which photon propagation and gravitational lensing can be used to probe the curvature structure of wormhole spacetimes and distinguish them from black holes.

\bibliography{references}

\end{document}